\documentstyle[aps,preprint]{revtex}
\tighten

\newcommand{\beq}{\begin{equation}}
\newcommand{\eeq}{\end{equation}}
\newcommand{\bea}{\begin{eqnarray}}
\newcommand{\eea}{\end{eqnarray}}
\newcommand{\simg}{\gtrsim}
\newcommand{\siml}{\lesssim}
\newcommand{\lmk}{\left(}
\newcommand{\rmk}{\right)}
\newcommand{\lkk}{\left[}
\newcommand{\rkk}{\right]}

\begin{document}

\draft
\preprint{astro-ph/00xxxxx, KNGU-INFO-PH-15}

\title{Enhanced black hole formation by multiple winding cosmic strings}

\author{Michiyasu Nagasawa}

\address{Department of Information Science, Faculty of Science,
Kanagawa University, Kanagawa 259-1293, Japan \\
Electronic address: nagasawa@info.kanagawa-u.ac.jp}

\maketitle

\begin{abstract}

Primordial black hole formation by cosmic string collapse is reconsidered
when the winding number of the string is larger than the unity.
The line energy density of the multiple winding string becomes greater than
that of the single winding string so that the probability of the black hole
formation by the string collapse during the loop oscillation would be strongly
enhanced. Although the production of the multiple winding defect is suppressed
and its number density should be small, the enhancement of the black hole
formation by the increased energy density may provide a large number of
evaporating black holes in the present universe which gives more stringent
constraint on the string model compared to the ordinary string scenario.

\end{abstract}


\section{Introduction}

Topological defects\cite{KVS} are considered to be produced in the early
universe during the cosmological phase transitions accompanied by some kinds
of symmetry breaking. They would contribute not only to the experimental
investigation of particle physics models in our universe as a high energy
laboratory but also to the theoretical explanation of various unresolved
problems in the standard Big-Bang cosmology.
Here we consider the primordial black hole formation by cosmic strings
which is one of the notable subjects both in the observational astrophysics
and the modern particle cosmology.

A cosmic string is a line-like topological defect which can be characterized
almost by one model parameter, $\eta$, the symmetry breaking scale of the
phase transition in which the string is generated. Particularly the line
energy density of the string, $\mu$, is expressed as
\beq
\mu \simeq \eta^2\ ,
\eeq
using $\eta$. In the course of the cosmic evolution, the size and number
distribution of the string is believed to obey the scaling model.
In the scaling distribution, the ratio of the energy density of the cosmic
string to the cosmological background energy density should be time
independent. The numerical value of this ratio can be written approximately by
\beq
G\mu =\lmk \frac{\eta}{M_{pl}}\rmk^2\ ,
\eeq
where $M_{pl}$ is the Planck mass.
Note that when $\eta \sim$ the grand unification scale, that is, $10^{15-16}$
GeV,
\beq
G\mu \sim 10^{-6}\ ,
\eeq
which is suitable for the initial seed
amplitude of the cosmological structure formation. This is why the GUT scale
string has been not only regarded as one of the possible consequence from some
unified theories but also considered to be a promising candidate of
the required density perturbation seed for galaxies and clusters of galaxies.
The viability of such a scenario can be investigated by the direct comparison
between the results of numerical simulations and the various observational
facts such as the galaxy distribution and the cosmic microwave background
radiation fluctuation similarly to other cosmological structure formation
scenarios. Although in the literature, the string scenario seems to be a less
promising one than that based on the inflationary epoch, the fact that there
is no established version of cosmological structure formation scenario
suggests cosmological defects may have played an important role in the cosmic
evolution history.

In this paper, we pay particular attention to the constraint on the model
parameter in which the cosmic string is produced using the black hole
formation by loop oscillation. The point which has not been discussed so far
is we employ the multiple winding number string as black hole sources.
In the next section and the third section, we briefly review the primordial
black hole formation by cosmic strings and the production of multiple winding
strings. Based on the results obtained in these two sections, the revised
observational constraint on the string model is shown in the section 4.
The final section is devoted to discussions and conclusions.

\section{PBHs from Cosmic Strings}

In this section, let we briefly summarize the primordial black hole formation
from oscillating string loops.
Black holes are one of the most exciting consequence of the general
relativistic gravitation theory and their astrophysical and cosmological
influences have been investigated by many people. Although the conventional
source of their production is the gravitational collapse of astronomical
bodies, the primordial formation in the early universe is also an interesting
and important issue which is worth to be studied in detail.

Particularly black holes whose mass is $M_*\simeq 10^{15}$g have the life-time
which is comparable to the age of the universe so that they would evaporate
at present and may provide sources of ultra high energy cosmic rays
including extragalactic $\gamma$-rays\cite{gamma} and other astroparticle
phenomena whose mechanisms have not been solved completely. In addition to
that, the relic black holes might contribute to the energy density of
the present universe and solve the dark matter problem which is one of
the most important cosmological problems\cite{MacG}. Such primordial black
hole formation could be predicted not only by the density perturbations
produced during the inflationary expansion of the universe but also by
the collapses of oscillating string loops.

In the standard picture of the cosmic string evolution, it is believed that
the distribution of the size and number of strings obeys the one scale model,
that is, the scaling distribution. In this formula, the string can be
classified to two kinds of string, one is an infinitely long horizon-scale
string and the other is a loop. As we mentioned in the previous section,
the relative energy density of infinite strings is scale-invariant since
they lose their energy in the course of cosmic evolution. In the conventional
model, the main channel of the infinite string energy loss is considered to be
a loop generation by the small-scale structure of long strings. Although there
exists a recent claim which says that the infinitely long strings have very
small wiggles on them so that the dominant energy loss mechanism of
the string should be a direct particle production\cite{VAH}, the conclusion
that the resulting string distribution can be described by the scaling formula
is unchanged. In this formula, the energy density of the infinite string can
be written as
\beq
\rho_{\infty}=\frac{\mu}{L^2}\ , \label{eq:rhoinf}
\eeq
where $L$ is the characteristic length scale of the string network which can
be defined as
\beq
L=\lmk \zeta \rmk^{-1/2}t\ ,
\eeq
using the parameter, $\zeta$ which is a typical number of infinite strings
within the horizon. On the other hand, the loop number density $n(l,t)$
for the loop size, $l$, and the time, $t$, can be expressed as
\beq
n\lmk l,t \rmk =\frac{\nu}{t^{3/2}\lmk l+\Gamma G\mu t\rmk^{5/2}}\ ,
\eeq
where $\nu$ is related to the loop creation length, $\gamma t$,
and $\zeta$ as
\beq
\nu \simeq \zeta\sqrt{\gamma}\ ,
\eeq
and $\Gamma$ is the efficiency of the gravitational wave radiation.
If other way of energy emission is dominant, $\Gamma G\mu$ should be
replaced appropriately.

Whether the energy loss mechanism of the string loop is the gravitational
wave radiation or not, oscillating loops may collapse to black holes.
When the loop whose total length is equal to $l$ is compactified into
the region whose radius is $2G\mu l$ during its oscillation, all the energy
the loop has is contained within the gravitational radius so that a black
hole of mass $\mu l$ would appear. Although such a procedure still remains
a conjecture, it is believed that the probability of black hole formation
during a loop oscillation, $P_{BH}^1$, can be written as
\beq
P_{BH}^1\sim \kappa \lmk G\mu\rmk^{\alpha}\ , \label{eq:probBH}
\eeq
where $\kappa$ is a numerical coefficient and $\alpha$ is a power index
which have been estimated as\cite{Hawking}
\beq
\alpha \simeq 0\sim 4\ ,
\eeq
or\cite{PZ}
\beq
\alpha \simeq -0.5\sim 0.5\ ,
\eeq
and\cite{CC}
\bea
\alpha &=& 4.1\pm0.1\ ,\\
\kappa &=& 10^{4.9\pm 0.2}\ ,
\eea
based on various assumptions and calculations.
Hereafter for simplicity, we employ a numerical value $\alpha =4$ as
the most distinct case. Since $\kappa$ is cancelled out in the calculation
process, its value does not affect the final conclusion.

Combined with the observational constraints obtained from the existence
of primordial black holes, the upper limit on the important string parameter,
$G\mu$, can be derived. Although the evolution and the final destiny of
the black hole have not been completely determined yet, there are two kinds
of constraints by the evaporation of the black hole and its massive relics,
which are naturally deduced from the assumption that the black hole loses
its energy and mass due to the Hawking radiation.
The former constraint is obtained from the observation of galactic and
extra-galactic cosmic rays such as $\gamma$-ray bursts, $\gamma$-ray
background, anti-protons and so on. The latter is based on the consideration
that the black hole should not disappear away but leave a massive relic.
Then it would contribute to the energy density of the universe and the total
mass can be limited by the fact that the cosmic evolution must obey
the standard history and the universe must not be over-closed.

Although there are many uncertainties such as what the final state of
the black hole is, how much the loop number density is and how much
the black hole formation probability is, the summarized constraint results
in\cite{MBW}
\beq
G\mu \siml \lmk 1-3\rmk \times 10^{-6}\ . \label{eq:Gmu}
\eeq
Note that this is in rough agreement with the bound by the cosmic microwave
background radiation anisotropy as\cite{CMB}
\beq
G\mu \siml \lmk 1-2\rmk \times 10^{-6}\ ,
\eeq
and the most stringent constraint on the string model by the pulsar timing
observation\cite{Pulsar}. As we mentioned in the previous section, the above
conditions become marginal when $\eta \sim$ the GUT scale.

\section{Multiple Winding String}

In the literature, only the simplest string configuration, that is, a string
whose winding number, $n$, equals one was investigated since it was implicitly
assumed that it seemed to be natural. Recently, however, it is claimed that
the multiple winding string may play an important role in the cosmological
evolution of the early universe. One is the initial condition for
the inflationary universe and the other is the electroweak baryogenesis.

The inflation scenario is the most promising paradigm which can solve
many problems the standard Big-Bang theory cannot explain.
So far, however, there has been no convincing model of the inflation
and people have invented many models. Among them, the topological
inflation\cite{VL} model can naturally provide an initial condition for
the inflationary expansion since at the core of the topological defect
the inflaton energy takes large amplitude which enables the dominance of
the effectively constant energy density. The difficulty of this efficient
model is that in order to satisfy the condition that the whole universe within
the horizon must be dominated by the inflaton field, the symmetry breaking
scale of the defect forming phase transition should be sufficiently large as
\beq
\eta \simg M_{pl}\ .
\eeq
Since no one knows a reliable physical theory at such a high energy scale,
it would be useful if this constraint can be lowered.
In the case of the topological string, the multiple winding number can
settle the situation since the larger the winding number becomes, the larger
the core size of the string grows, which results that the required value of
$\eta$ decreases. Particularly when
\beq
n\cong 3\times 10^3\ ,
\eeq
the topological inflation can occur at the GUT scale\cite{LTV}.

The baryon asymmetry problem is one of the most significant problems in
modern cosmology and particle physics. Observations show the fact that
there exists small baryon number as
\beq
\frac{n_B}{s}\sim 10^{-10}\ ,
\eeq
which is the ratio of the baryon number density to the entropy.
To explain the generation of baryon asymmetry in the course of cosmic
evolution, three necessary conditions are required. These are the baryon
number violation, C and CP violation and the deviation from the thermal
equilibrium. The first and the second conditions may be satisfied by
the appropriate model construction of particle interactions.
On the contrary, the last one should be provided during the dynamical
evolution of the universe.

Because of the sphaleron transition process, all the baryon asymmetry
should be erased at the electroweak epoch unless there exists an asymmetry
between baryon number and lepton number. This is the reason why
the electroweak baryogenesis has been regarded as the most promising
scenario of baryon number generation for many years\cite{Trodden}.
In the conventional model of the electroweak baryogenesis, the deviation
from the thermal equilibrium is achieved by the propagation of nucleated
bubbles and the interaction between bubble walls and the surrounding plasma
during the first order electroweak phase transition. However, in the standard
model and its simple extension, the degree of the electroweak phase transition
seems to be too weak to realize the first order transition.

In such a situation, various alternatives
have been suggested and the electroweak string has been proved to bring
the non-equilibrium condition even if the electroweak phase transition is
not of first order since it should be left after the transition and its
collapse resembles the bubble wall propagation\cite{BD}. Although the idea
that the defect can realize the out-of-equilibrium remains valid, it has
been shown that the initial production rate of the electroweak string in
the standard model is too low so that it is difficult to explain
the observational value quantitatively by this scenario\cite{NY}.
Recently more effective scenario using the string has been proposed\cite{Soni}.
In this case, when
\beq
n\geq 2\sim 3\ ,
\eeq
the deviation from the thermal equilibrium can be satisfied by
sphaleron bound states on strings and their following decay.

Therefore it would be useful if the existence of the multiple winding
string can be constrained by some astronomical and cosmological observations
and here we consider the primordial black holes produced by the string loop
oscillation. In order to perform a quantitative analysis, we estimate
the formation probability of the multiple winding string.
Although similar consideration can be applied to the multiple winding
monopole which may be useful for the topological inflation, hereafter we
will concentrate on the string case.

We employ the Abelian Higgs model with the Lagrangian as
\bea
{\cal L} &=& -\frac{1}{4}F^{\mu\nu}F_{\mu\nu}
-\frac{1}{2}(D^{\mu}\phi)^{\dagger}(D_{\mu}\phi)
-\frac{1}{8}\lambda({\phi}^{\dagger}\phi - {\eta}^2)^2\ ,\\
& & D_{\mu}={\partial}_{\mu}-ieA_{\mu}\ ,\\
& & F_{\mu\nu}={\partial}_{\mu}A_{\nu}-{\partial}_{\nu}A_{\mu}\ ,
\eea
where $\phi$ is a complex scalar field, $A_{\mu}$ is a gauge vector field,
$e$ is the gauge coupling constant.
In this model, there is a well-known string solution called
the Nielsen-Olesen vortex\cite{NO} which is a two-dimensional slice of
the string and the string configuration with a fixed winding number, $n$,
can be determined by one parameter, $\beta$, which can be defined as
\beq
\beta =\frac{\lambda}{e^2}=\lmk \frac{m_s}{m_v}\rmk^2\ ,
\eeq
where $m_s$ is the mass of the scalar field and $m_v$ is the mass of
the vector field. The important characteristic of the string configuration is
that as $\beta$ or the winding number, $n$, increases, the width scale of
the string core also increases.

The stability of the multiple winding string has been analyzed and 
in some parameter range, it is stable. For example, the calculation of
the string line energy density shows that the multiple winding string is
stable for the case $\beta <1$ and unstable for $\beta >1$\cite{BV}.
Moreover, the interaction force between two vortices is attractive when
$\beta <1$ and repulsive when $\beta >1$\cite{JRBR}.

Now we estimate the formation probability of the multiple winding string.
In this paper, we review the calculation process briefly and the detailed
method can be found in the reference\cite{ON}.

Before introducing how to estimate the string formation probability for
the multiple winding, let us summarize that for the single winding
case\cite{VVP}. In the Kibble mechanism, the phase of the Higgs field is
considered to take random value at each correlated region whose size is of
scale, $\xi$, which is the correlation length of the fields. The result shows
that the formation probability of the single winding string, $P(1)$ is 
\beq
P(1)=\frac{\pi}{2}/2\pi =\frac{1}{4}\ .
\eeq
In this procedure, the string identification process can be described as
follows. First divide the physical space into domains whose size is typically
of $\xi$, that is, the correlation scale. Then assign the phase of the Higgs
field randomly to each representative point of every domain. Finally
interpolate the phase between two representative points so that the gradient
energy of the Higgs field takes the minimum value. Thus we can distinguish
the region where there is a winding number and a string exists from that
where there is no winding number and we cannot find a string.

Based on the above argument let us proceed to the multiple winding string
version. If we simply apply the Kibble mechanism procedure to the multiple
winding case, we have to encounter a problem. Since the phase difference
between two neighboring representative points in the triangle division method
which we have described, the total difference of the phase along the triangle
circumference should be smaller than or equal to $3\pi$, which means it must
be $2\pi$ at most because of the phase continuity. Thus the winding number
cannot be larger than the unity. The most natural solution to this situation
may be the employment of other polygon than the triangle, for example,
a hexagon. In the hexagon case, the total phase difference is improved to
be $6\pi$ and the region within which the winding number becomes two can be
allowed. However, there remains a problem even in this case because we cannot
distinguish the state there is one piece of double winding string from
the state there are two pieces of single winding string. Moreover, if we
apply the triangle method to the area where one double winding string can
be found by the hexagon method, we will misunderstand that there is
one single winding string. Therefore not only the winding number but also
even the magnitude of the total winding depends on the method of space
dividing.

Thus in order to estimate the formation probability of the multiple winding
string correctly, we introduce a new scale, $R$, within which there should
be only one piece of string. Then the number of vertices the polygon has
becomes $\pi R/\xi$. Thus the total phase difference is
\beq
\pi \cdot \frac{\pi R}{\xi}=\frac{\pi^2 R}{\xi}\ .
\eeq
The most natural interpretation would be that $R$ is comparable to
the diameter of the string core which is approximately equal to the inverse
of the scalar mass which results
\beq
R= \frac{1}{m_s(T=T_G)} \sim \frac{1}{\lambda T_G}\ ,
\eeq
where $T_G$ is the temperature at which the phase transition terminates and
the configuration of topological defects cannot be erased by the thermal
fluctuations of fields. On the other hand, the correlation scale, $\xi$,
can be expressed as\cite{YY}
\beq
\xi \sim \frac{1}{T_G}\ .
\eeq
As a consequence, the possible maximum winding number which belongs to
one string should be
\beq
n_{max} \sim \Bigg[\frac{\pi^2 R}{2\pi\xi}\Bigg] \sim
\Bigg[\frac{\pi}{2\lambda}\Bigg]\ , 
\eeq
where $[\ ]$ denotes the Gauss's symbol.
Thus we can say that the multiple winding string can be produced
when $\lambda <<1$.

Even in the case $\lambda \sim 1$, we cannot completely deny the formation
of the multiple winding configuration since in the actual situation,
the correlation length scale can be inhomogeneous, that is, be various at
each correlated region. Based on the result of the numerical calculation
using the toy model\cite{ON}, we estimate the black hole formation possibility
in the next section.

Before closing this section, let us mention another possibility.
When the self-coupling constant of the scalar field is smaller
than the gauge coupling constant, that is, $\beta <1$ for the local
gauged string case, the attractive force acts on strings and they
would accumulate to a multiple winding string. In this case,
the formation probability of the string whose winding number equals $n$,
$P(n)$, can be easily calculated as
\beq
P(n)=\frac{P(1)}{n}\ . \label{eq:Pn}
\eeq
This is also the case which the black hole formation probability is
calculated in the following section.

\section{PBHs from Multiple Winding Loops}

In this section, we estimate the primordial black hole formation probability
by the oscillating string loop and the resulting modified observational
constraint. Since we have the formula for the ordinary single winding string
case in (\ref{eq:Gmu}), the only thing we have to do is to clarify
the difference when we employ the multiple winding one.
The constraint (\ref{eq:Gmu}) was deduced by the upper bound on $\Omega_{BH}$,
the relative energy density of black holes to the critical energy density
of the present universe. Hence in the following we will derive that for
the $n$ winding string, $\Omega_{BH}^n$.

Here we assume the modified formula of the parameter which characterizes
the multiple winding string. If the line energy density of $n$ winding
string, $\mu_n$, can be written as
\beq
\mu_n =A(n)\mu\ , \label{eq:Anmu}
\eeq
using the enhancement factor, $A(n)>1$, then
\beq
A(n)\leq n\ ,
\eeq
since otherwise a multiple winding string becomes energetically unstable to
the division into $n$ pieces of single winding strings.
As we have introduced in the equation (\ref{eq:probBH}), the formation
probability of the black hole by the loop oscillation is in proportion to
the power of the string line energy density so that this probability
can be enhanced as
\beq
\frac{P_{BH}^n}{P_{BH}^1}=\lkk \frac{A(n)}{A(1)}\rkk^{\alpha}
=A(n)^\alpha\equiv f_1\ ,
\label{eq:fone}
\eeq
where the superscript of $P_{BH}$ depicts the corresponding string winding
number and we define this enhancement factor as $f_1$.

In order to see how the constraint on the string model should be replaced,
it is also necessary to estimate the seed loop number density
in addition to the above black hole formation probability.
Using the assumption in the equation (\ref{eq:Anmu}) and one more assumption
that the multiple winding string also obeys the scaling distribution
described by the formula (\ref{eq:rhoinf}), we can calculate
explicitly two modifications by the increase of the winding number which
affect the black hole number density at present. One is the reduction
factor which comes simply from the formation probability of multiple winding
strings and the other is the enhancement factor which is related to earlier
formation of black holes.

First the energy density of $n$ winding strings, $\rho_{\infty}^n$, is
smaller than that of single winding strings, $\rho_{\infty}$, as
\beq
\rho_{\infty}^n=P(n)\frac{\mu_n}{L^2}\ , \label{eq:rhoinfn}
\eeq
where we have written the probability of $n$ winding string production
per correlation volume at the initial formation epoch as $P(n)$ similarly
to the previous section. Then there appears a modification to the final
formula of the black hole density parameter as
\beq
\Omega_{BH}^n\propto P(n)\ ,
\eeq
which results an relative difference compared to a single winding case as
\beq
f_2\equiv \frac{P(n)}{P(1)}\ , \label{eq:ftwo}
\eeq
where we define this reduction factor as $f_2$.

The next point is the time, $t_{BH}^n$, that the black hole of specific
mass $M_*$ which is evaporating at present was formed becomes earlier in
the multiple winding string case. Since $t_{BH}^n$ can be calculated by
the relation as
\beq
M_*=\mu_n \gamma t_{BH}^n\ , \label{eq:Mstar}
\eeq
$t_{BH}^n$ becomes smaller as $\mu_n$ increases. Strictly speaking the value
of $M_*$ should depend on the time of formation. However, the rate of
change of the black hole mass, $M$, is proportional to the power of $M$
as\cite{PM}
\beq
\frac{dM}{dt}\propto M^{-2}\ .
\eeq
Then the life-time of the black hole is much larger than its creation
epoch. For example, $t_{BH}^1\sim 10^{-16}$ sec for the GUT scale string is
sufficiently smaller than the age of the universe, $t_0\sim 10^{17}$ sec.
Thus it is a good approximation that $M_*$ is independent of the string
winding number, $n$, so that we can say that
\beq
t_{BH}^n\propto \frac{1}{\mu_n}\ . \label{eq:tnBH}
\eeq
One more note is that in the equation (\ref{eq:Mstar}) it is assumed that
the black hole is formed instantaneously after the loop creation.
However, since the important feature is the relation between $t_{BH}^n$
and $\mu_n$, we do not investigate whether this is appropriate or not
in detail.

In order to calculate $\Omega_{BH}^n$, we have to take various mass of
the black hole into account. Hereafter, however, we only estimate the
number density of the black hole whose mass is $M_*$. This simple
consideration is sufficient to see the essential dependence on $\mu_n$
of $\Omega_{BH}^n$ since the black hole of mass $M_*$ has the most dominant
effect and the exact treatment produces the completely same result.
The initial black hole number density whose mass equals $M_*$,
$n_{BH}^n(t_{BH}^n)$, can be calculated as
\beq
n_{BH}^n(t_{BH}^n)=P_{BH}^n n_{loop}^n(t_{BH}^n)\ ,
\eeq
where $n_{loop}^n$ is the loop number density of the $n$ winding string and
can be written as
\beq
n_{loop}^n(t_{BH}^n)\simeq \frac{\Delta\rho_{\infty}^n(t_{BH}^n)}{M_*}\ ,
\eeq
where $\Delta\rho_{\infty}^n$ is the energy loss of an infinitely long $n$
winding string to loops. Using the formula (\ref{eq:rhoinfn}),
\beq
\Delta\rho_{\infty}^n(t_{BH}^n)\simeq C\rho_{\infty}^n(t_{BH}^n)
=CP(n)\zeta\frac{\mu_n}{\lmk t_{BH}^n\rmk^2}\ ,
\eeq
where $C$ is the efficiency of the loop production. When we leave only
the dependence on the winding number, 
\beq
n_{BH}^n(t_{BH}^n)\propto P_{BH}^nP(n)\frac{\mu_n}{\lmk t_{BH}^n\rmk^2}\ .
\label{eq:nBH}
\eeq

The following evolution of the black hole number density, $n_{BH}^n$, obeys
the ordinary dust particle law as
\beq
n_{BH}^n(t) \propto a(t)^{-3}\ ,
\eeq
where $a(t)$ is the cosmological scale factor. Then since $t_{BH}^n$ is in
the radiation dominated era for the parameter region we are interested in,
the present number density of the black hole whose mass is equal to $M_*$
can be expressed as
\bea
n_{BH}^n(t_0) &=& n_{BH}^n(t_{BH}^n)\lkk\frac{a(t_{BH}^n)}{a(t_0)}
\rkk^3 \nonumber \\
&=& n_{BH}^n(t_{BH}^n)\lmk\frac{t_{BH}^n}{t_{eq}}\rmk^{3/2}
\lmk\frac{t_{eq}}{t_0}\rmk^2 ,
\eea
in the flat universe where $t_{eq}$ is the radiation matter equality time.
Then the density parameter for the black hole can be written as
\beq
\Omega_{BH}^n=\frac{M_*n_{BH}^n(t_0)}{\rho_0}\ ,
\eeq
if we neglect the mass loss of the black hole where $\rho_0$ is the critical
density of the universe. Combined with the relations (\ref{eq:nBH}) and
(\ref{eq:tnBH}), we can obtain the final relation as
\beq
\Omega_{BH}^n\propto P_{BH}^nP(n)\mu_n^{3/2}\ ,
\eeq
and we define an additional modification factor, $f_3$, as
\beq
f_3\equiv \lmk\frac{\mu_n}{\mu}\rmk^{3/2}=\lkk\frac{A(n)}{A(1)}\rkk^{3/2}
=A(n)^{3/2}\ . \label{eq:fthree}
\eeq

As a result, the overall modification factor of $\Omega_{BH}^n$ for the string
whose winding number is equal to $n$, $f(n)$, should be described as
\beq
f(n)=f_1f_2f_3=\frac{P(n)}{P(1)}A(n)^{\alpha+3/2}\ ,
\eeq
using the equations (\ref{eq:fone}), (\ref{eq:ftwo}) and (\ref{eq:fthree}).
Thus when $A(n)>1$ at a certain $n$, this enhancement may overcome
the reduction by the smallness of $P(n)$ and make the present black hole
number larger compared to the single winding string case although the loop
creation size and other parameters are not completely determined in
the multiple winding string case.

At last we can estimate the modification factor of the density parameter
of the black hole at present for the multiple winding string. The first case
is the string formation probability is calculated including the fluctuation
of the correlated region size. The assumption for the parameter is
\beq
\alpha =4\ ,\quad A(n)=n\ ,
\eeq
which makes the modification rather large. The final expression of
the factor, $f(n)$, can be written as
\beq
f(n)=\frac{P(n)}{P(1)}n^{5.5}\ ,
\eeq
and the numerical values for $n=1-4$ are depicted in Table 1.
In this case, $f_1$ and $f_3$ is too small to enhance the black hole
production so that the result is trivial, that is, the constraint on
the string model parameter is unchanged since the most dominant contribution
to the $\gamma-$ray emission and other astrophysical phenomena would be
provided by single winding strings. Although we do not know the numerical
value of $P(n)$ for much larger $n$, it would be natural that $f(n)$ is
the decreasing function of $n$ even in such cases.

\begin{center}
\begin{tabular}{|c|c|c|}
\hline
$n$ & $P(n)$ & $f(n)$ \\
\hline
$1$ & $0.2102$ & $1$ \\
$2$ & $8.36\times 10^{-4}$ & $1.80\times 10^{-1}$ \\
$3$ & $4.8\times 10^{-7}$ & $9.6\times 10^{-4}$ \\
$4$ & $8\times 10^{-11}$ & $8\times 10^{-7}$ \\
\hline
\end{tabular}
\vspace{0.5cm}

{\bf Table 1.\quad Modification Factor for Multiple Winding String}
\end{center}

In contrast to that, the other case in which the string coalescence is
taken into account may be more interesting. Using the same assumption for
the parameters, $\alpha$ and $A(n)$, as the former case, the formula of
the modification factor can be calculated as
\beq
f(n)=n^{4.5}\ ,
\eeq
where we have substituted the equation (\ref{eq:Pn}). The factor $f(n)$ are
shown in Table 2 for $n=2-4$. It can be obviously seen that
the degree of the enhancement becomes greater as $n$ increases.
Therefore it may be plausible that the string which has a large number of
winding might impose a more stringent constraint than the single winding
string scenario.

\begin{center}
\begin{tabular}{|c|c|c|c|}
\hline
$n$ & $2$ & $3$ & $4$ \\
\hline
$f(n)$ & $22.6$ & $140$ & $512$ \\
\hline
\end{tabular}
\vspace{0.5cm}

{\bf Table 2.\quad Modification Factor for String Coalescence Case}
\end{center}

Finally we show the modified constraint formula for the string model
parameter, $\eta$, which is significant from the point of view of the particle
cosmology. As we have seen in this section, the symmetry breaking parameter
is related to the black hole number as
\beq
\Omega_{BH}^n\propto \mu_n^{\alpha+3/2}\ .
\eeq
Then the revised upper bound on $G\eta^2$ for the multiple winding string
case, $\left. G\eta^2\right|_<^n$, can be expressed as
\beq
\left. G\eta^2\right|_<^n=f(n)^{-2/\lmk 3+2\alpha \rmk}\left.
G\eta^2\right|_<\ ,
\eeq
where $\left. G\eta^2\right|_<$ is the upper bound on $G\eta^2$ for
the case in which the multiple winding string is not incorporated.

\section{Discussion and Conclusion}

In this paper, we have estimated how the observational constraint on
the particle physics model parameter in which the cosmic string is
produced should be modified when we consider the multiple winding string
which may be useful for the topological inflation and the electroweak
baryogenesis. The modified formula of the constraint on the line energy
density which can be translated to that on the symmetry breaking scale,
$\eta$, can be written as
\beq
G\eta^2 \siml f(n)^{-2/\lmk 3+2\alpha \rmk}\times 10^{-6}\ ,
\eeq
in which we assume the simple scaling distribution of the string.

The numerical values of the above modification factor are calculated
in two cases. One is the case that the multiple winding string is produced
at the initial formation epoch and the modification is not so significant
since
\beq
f(n)\leq 1\ .
\eeq
In the other case, the dynamical evolution of the string after its formation
is concerned and the string accumulation by the attractive interaction
between strings makes the enhancement remarkable as
\beq
f(n)>>1\ ,
\eeq
which means the upper bound on $\eta$ must be much lower.

Thus we can say that a more stringent constraint by primordial black holes
on the model parameter might be obtained by multiple winding strings compared
to the conventional scenario of ordinary single winding strings.
Just for the reference, let us estimate the case in which all the winding
number within the horizon are concentrated on a single string.
At the string formation epoch, the number of the correlated region in
the horizon volume can be estimated as
\beq
N\cong \lmk\frac{t}{T^{-1}}\rmk^3\ .
\eeq
When we take the annihilation of a string by an anti-string into consideration,
the bulk total winding within the horizon would be
\beq
n\cong \sqrt{P(1)N}\ .
\eeq
If we assume $T\sim \eta$, then
\beq
n\simeq \lmk \frac{\sqrt{\cal N}}{0.3P(1)^3}\frac{\eta}{M_{pl}}\rmk^{-3/2}\ ,
\eeq
using the relation between $t$ and $T$ in the standard cosmology where
${\cal N}$ is the gauge degree of freedom of particles at $T$.
Then the modified constraint on $\eta$ can be written as
\beq
\lmk \frac{\eta}{M_{pl}}\rmk^{1/2+6/(2\alpha +3)} \siml \lmk
\frac{\sqrt{\cal N}}{0.3P(1)^3}\rmk^{3(2\alpha -1)/2(2\alpha +3)}10^{-6}\ ,
\eeq
leaving the parameter $\alpha$ in the formula (\ref{eq:probBH}).
When $\alpha =4$, the resulting constraint can be expressed as
\beq
\frac{\eta}{M_{pl}}\siml 10^{-4}\ ,
\eeq
which is marginally compatible with the GUT string, $\eta\simeq 10^{15}$
GeV and a little more rigorous than the original one (\ref{eq:Gmu}).
It might be the most useful condition in order to constrain the model
construction of the string generation in the early universe even if
the cosmic microwave background radiation fluctuation and the pulsar timing
are included.

This is not the end of the story. There may be other effects which can enable
the abundant formation of the multiple winding defect such as the relaxation
of the geodesic rule\cite{PV} which is applied when the phase between two
different correlated regions is interpolated.
Moreover, it may be possible
that the homogeneous gauge flux distribution greatly enhances the total
winding number within the correlated domain so that huge winding strings
might be produced by the string coalescence\cite{HR}.
Further work would be needed in the future.

\acknowledgments

The author is grateful to Arttu Rajantie for his comment at the international
workshop on Particle Physics and The Early Universe, COSMO2000.

\end{document}